\documentclass[aps,prd,showkeys,preprint,12pt]{revtex4-1}
\usepackage[english]{babel}
\usepackage{amsmath,amsthm,amsfonts,amssymb}
\usepackage{dsfont}
\usepackage{graphicx}
\usepackage{caption}
\usepackage[mathcal]{eucal}
\usepackage{mathrsfs}
\usepackage{graphicx}
 \usepackage{graphics}
\usepackage{dcolumn}
\usepackage{latexsym}
\usepackage{epsfig}
\usepackage{bm}
\usepackage[all]{xy}
\usepackage[T1]{fontenc}
\usepackage{xcolor}

\graphicspath{{./img/}}

\newcommand{\ie}{\begin{equation}}
\newcommand{\fe}{\end{equation}}
\newcommand{\se}{\begin{eqnarray}}
\newcommand{\ff}{\end{eqnarray}}

\begin{document}

\title{Thermodynamics of Schwarzschild-like black holes in modified gravity models}

\author{D. A. Gomes}
\email{deboragomes@fisica.ufc.br}

\author{R. V. Maluf}
\email{r.v.maluf@fisica.ufc.br }

\author{C. A. S. Almeida}
\email{carlos@fisica.ufc.br}

\affiliation{Universidade Federal do Cear\'a (UFC), Departamento de F\'isica, Campus do Pici, Fortaleza - CE, 60455-760 - Brazil}

\begin{abstract}
Over the last decades, many methods were developed to prove Hawking radiation. Recently, a semiclassical method known as the tunneling method has been proposed as a more straightforward way of derivating black hole thermodynamical properties. This method has been widely applied to a vast sort of spacetimes with satisfactory results. In this work, we obtain the black hole thermodynamics in the presence of a Lorentz symmetry breaking (LSB). We apply the Hamilton-Jacobi method to Schwarzschild-like black holes, and we investigate whether the LSB affects their thermodynamics. The results found show that the LSB not only changes the black hole thermodynamic quantities but also makes it necessary to modify the standard first law of thermodynamics.\end{abstract}




\maketitle

\section{Introduction\label{sec:introduction}}
Although a black hole is classically defined as an object that can only absorb radiation, when it comes to a quantum mechanical approach, it actually emits radiation. This intriguing result was shown by Hawking in 1975 using quantum field theory in curved space-time \cite{Hawking}. After that, black holes assumed an essential role in the attempt of constructing a quantum theory of gravity. 

The concept of the so-called Hawking radiation puts the thermodynamical description of black holes in a more realistic basis. As a matter of fact, black hole mechanical properties were earlier discussed by Bardeen, Carter, and Hawking (1973) \cite{bardeen}. In this work they formulated the four laws of black hole mechanics, after the ideas of Bekenstein (1972) \cite{bekenstein1973}. These laws were proposed due to the close similarity of some mechanical properties of black hole to their entropy and temperature. Nowadays, black holes can be considered as thermal systems, and this set of laws are called the four laws of black hole thermodynamics. 

The Lorentz Symmetry is one of the most important and tested symmetries in physics, especially for General Relativity and the Standard Model. However, since it is expected that this symmetry should be spontaneously broken at the Planck scale, it is interesting to investigate the physical effects arising from this violation. On gravitational grounds, it is known that the Lorentz Symmetry Breaking (LSB) can modify the geometry of a black hole (for instance, see \cite{bertolami,casana,magueijo}, so it is natural to ask if black hole mechanical laws hold in LSB scenarios.  

Recently, it was argued that Lorentz invariance breaking might lead to the contradiction of the generalized second law of black hole thermodynamics \cite{eling,jacobson}. In fact, the generalized second law does not hold in LSB theories where different fields propagate at different speeds. This happens because, in these theories, the concept of horizons is field-dependent, and thus, the definition of the black hole entropy is ambiguous \cite{dubovsky}. However, this problem does not apply to every LSB theory \cite{shenji,betschart,benkel} and, indeed, it may be caused by the presence of negative energy configurations \cite{feldstein}. 

Therefore, the main purpose of this work is to study the thermodynamic properties of Schwarzschild-like black hole in scenarios in which the so-called Lorentz symmetry breaking (LSB) occurs.

A recent example of LSB applied to modify black hole thermodynamics is found in the work of Li et al. \cite{Li}, where the authors changed the thermodynamic quantities directly. In the present work, the tunneling method is applied to LSB modified space-times, and we analyze how the LSB modifies the black hole thermodynamics. 

This paper is organized as follows. In Sec. \ref{LSB-bumblebee}, we briefly review the LSB in the context of bumblebee gravity. In Sec. \ref{tunneling}, the quantum tunneling method is also reviewed, and its equivalence to Hawking's method is confirmed. Then, we present our results for thermodynamical functions of the Schwarzchild black hole in two different kinds of Lorentz violating gravity model in Sec. \ref{casana} and Sec. \ref{sec:5}. The difference between both approaches is the kind of the metric proposed. For the sake of simplicity, we name B-metric the proposal of Bertolami and Paramos \cite{bertolami}, and we call C-metric the proposal of Casana et al.  \cite{casana}. Finally, we present further discussions and our conclusions in Sec. \ref{Conclusion}.
 
\section{\textbf{Lorentz Symmetry Breaking in a Bumblebee Gravity Model}}
\label{LSB-bumblebee}

In bumblebee models, the Lorentz Symmetry Breaking (LSB) occurs due the presence of dynamical terms of the vectorial field $B_\mu$, which is known as bumblebee field. The bumblebee action is given by \cite{bluhm} 
\begin{equation}
\mathcal{S} = \int d^4 x \, \mathcal{L}_{ B},
\end{equation}
where the bumblebee Lagrangian $\mathcal{L}$ is
\begin{equation} \label{eq:11}
\mathcal{L}_{ B} = \mathcal{L}_{ g} + \mathcal{L}_{ gB} + \mathcal{L}_{ K} + \mathcal{L}_{ V} + \mathcal{L}_{ J}.
\end{equation}
The terms in \eqref{eq:11} are, respectively: the gravitational Lagrangian, the gravity-bumblebee coupling Lagrangian, the dynamical Lagrangian of the field $B_\mu$, the Lagrangian which contains a potential $V$ responsible for the LSB, and the Lagrangian which contains the interaction terms of $B_\mu$ to the matter or other sections of the model.

The dynamical Lagrangian $\mathcal{L}_{ K}$ contains the kinetic terms of $B_\mu$, expressed through the field strength for $B_{ \mu}$, namely
\begin{equation}
B_{\mu \nu} = D_{\mu}B_{ \nu} - D_{\nu}B_{ \mu},
\end{equation}
where $D_\mu$ is the covariant derivative, defined according to the spacetime curvature.

The potential contained in $\mathcal{L}_{ V}$ is responsible for the LSB, and it is chosen to have a minimum in the bumblebee Vacuum Expectation Value (VEV), denoted by $b_{\mu}$. For this reason, we usually refer to $b_{\mu}$ as the LSB parameter. Thus, it must have the following functional form:
\begin{equation} \label{eq:12}
V = V(B^{\mu} B_{\mu} \pm b^2),
\end{equation}
where $b^2$, the norm squared of $b_{\mu}$, is a positive constant. The signs in \eqref{eq:12} determine whether the VEV $b_{\mu}$ is timelike or spacelike. If $V$ has a polynomial form, like $V(x) = \lambda x^2 /2$, the constant $\lambda$ is a coupling constant. The potential $V$ can also have the form $V(x) = \lambda x$ in which case $\lambda$ is a Lagrangian-multiplier field, responsible for the constraint $B^{\mu} B_{\mu} \pm b^2$; this form is interesting for sigma models since it preserves only the Nambu-Goldstone modes \cite{bluhm}.

The matter-bumblebee coupling term can be written as $\mathcal{L}_{M} = - e B_\mu J^{mu}_M$, where $J^{\mu}_M$ is the matter current. Sometimes, it is interesting to neglect the matter current and allow bumblebee-matter coupling to occurs only gravitationally in order to avoid possible stability issues in the theory \cite{bluhm2}. Therefore, the bumblebee action is given by
\begin{equation} 
S_B = \int d^4x \sqrt{-g} \left[\frac{1}{2 \kappa}( R + \xi  B^{\mu}B^{\nu}R_{\mu \nu}) - \frac{1}{4} B^{\mu \nu} B_{\mu \nu} - V(B^{\mu} B_{\mu} \pm b^2)
\right],
\end{equation}
where $\kappa = 8 \pi G$. The potential form is not relevant since it is assumed that the bumblebee field is fixed at its nonzero VEV, which makes $V=0$ and $V'=0$. The spacetime is considered to have no torsion, in such a way that the bumblebee field strength is given by $B_{\mu \nu} = \partial_{\mu}B_{ \nu} - \partial_{\nu}B_{ \mu}$. It is also considered that the spacetime is static and has spherical symmetry; thus, we can describe it with a Birkhoff metric $g_{\mu \nu} = diag(-e^{2 \phi}, e^{2 \rho}, r^2, r^2 \sin^2 \theta)$, where $\phi$ and $\rho$ are functions of $r$.

In this work, we will use the black hole solutions in LSB scenarios presented in Ref. \cite{bertolami,casana} and find their corresponding thermodynamical properties. In both works, the VEV $b_\mu$ is assumed to take the form $b_\mu = (0,b(r),0,0)$, with $b^2 = b^{\mu}b_{\mu} = const.$, which means that the LSB is purely radial.
In Ref. \cite{bertolami}, Bertolami and Paramos imposed the condition $\nabla_\mu b_{\mu} = 0$ in place of the usual prescription $\partial_\mu b_\mu = 0$. On the other hand, Casana 
\textit{et al.} \cite{casana} proposed that $b(r)$ has the explicit form $|b|e^{\rho}$, which yields to $\nabla_\mu b_{\mu} \neq 0$. Moreover, although the bumblebee-matter coupling term is not disregarded in this case, the matter sector energy-momentum tensor is imposed to be $T_{\mu \nu}^M =0$, as a requirement for obtaining vacuum solutions. As a result, the equation of motion found in \cite{casana} leads to the Schwarzschild-like metric 
\begin{equation} \label{eq:30}
ds^{2}=-\left( 1-\frac{2M}{r} \right) dt^{2} + (1 + \ell) \left( 1-\frac{2M}{r} \right)^{-1}dr^{2}+r^2d\Omega^{2},
\end{equation}
where $\ell = \xi b^2$, $M = G_N m$ is the geometrical mass and $d\Omega^{2}$ is the solid angle.

We can notice that, as in the case with no LSB, we have the singularities $r=2M$ and $r=0$. Now, we have to find out whether these singularities are physical or not. Thus, we have to calculate the Kretschmann scalar given by $K = R_{\mu \nu \alpha \beta} R^{\mu \nu \alpha \beta}$ for the metric \eqref{eq:30}, which is given as
\begin{eqnarray} \label{eq:21}
K_C &=& \frac{4(12 M^2 + 4 \ell M r + \ell^2 r^2)}{r^6(\ell + 1)^2} = \frac{K_S}{(\ell + 1)^2} + \frac{4(4 \ell M r + \ell^2 r^2)}{r^6(\ell + 1)^2},
\end{eqnarray}
where $K_S = 48 M^2 r^{-6}$ is the usual Kretschmann scalar for the Schwarzschild black hole with no LSB. We can notice from \eqref{eq:21} that, for $r= r_S = 2M$, the Kretschmann scalar is finite, which implies that this singularity is removable. For $r=0$, however, the Kretschmann scalar is infinite, which implies that this is a physical singularity. Therefore, the nature of the singularities is not modified by this radial LSB.


It is interesting to stress that, if we make a transformation $\tilde{r}=\sqrt{l+1}r$ in Eq.\eqref{eq:30}, we note that it becomes
\begin{equation}
    ds^2 = - \left( 1- \frac{2 \tilde{M}}{\tilde{r}} \right) dt^2 + \left( 1- \frac{2 \tilde{M}}{\tilde{r}} \right)^{-1} d\tilde{r}^2 + \frac{\tilde{r}^2}{1+\ell} d \Omega ^2,
\end{equation}
where we redefined the mass $M \rightarrow\tilde{M} = \sqrt{1+\ell}M$. As it is not possible to redefine the metric of the unit 2-sphere $d\Omega^2$, we can conclude that the C-metric does not reduce to the standard Schwarzschild metric.

As it was said before, Bertolami and Paramos impose $\nabla_\mu b_{\mu} = 0$, which yields, after some calculation, the following Schwarzschild-like metric
\begin{equation} \label{eq:14}
ds^{2}=-\left(1  - \frac{2 M}{r} \frac{r^L}{r_0^L} \right) dt^{2} + \left(1  - \frac{2 M}{r} \frac{r^L}{r_0^L} \right)^{-1}dr^{2} + r^2 d\Omega^{2},
\end{equation}
where $L = \xi \bar{b}^2/2$, $M = G_L m$ is the geometrical mass, $r_0$ is an arbitrary distance and $d\Omega^{2}$ is the solid angle. 

The event horizon is given by the condition $g_{0 0} = 0$, i.e.,
\begin{equation} \label{eq:13}
1  - \frac{2 M}{r} \frac{r^L}{r_0^L} = 0,
\end{equation}
which gives $r_B = (2 M r_0^{-L})^{1/(1-L)}$. The corresponding Kretschmann scalar is given by
\begin{eqnarray} \label{eq:15}
K_B &=& 48 \left[ 1 - \frac{5}{3} L + \frac{17}{12} L^2 - \frac{1}{2} L^3 + \frac{1}{12} L^4 \right] M^2 \left( \frac{r}{r_0} \right)^{2L} r^{-6} \nonumber \\
&\simeq &  \left( 1 - \frac{5}{3} L \right) \left( \frac{r}{r_0} \right)^{2L} K_S.
\end{eqnarray}
It can be seen from \eqref{eq:14} that $K(r = r_B)$ is finite and, therefore, this singularity is removable. Since $b$ is very small, $r^{2L-6} \rightarrow \infty$ as $r \rightarrow 0$, which means that the singularity $r = 0$ is intrinsic as in the usual case. We can notice that, unless $r_0 = 2M$, the nature of the singularities for the metric \eqref{eq:14} is modified.

Now, we can investigate the similarities between the metric \eqref{eq:14} and the so called Schwarzschild-Tangherlini metric \cite{tang}. For simplicity, we write the metric \eqref{eq:14} as
\begin{equation}
    ds^2 = - \left( 1- \frac{2 \bar{M}}{r^{1-L}} \right)dt^2 + \left( 1- \frac{2 \bar{M}}{r^{1-L}} \right)^{-1} dr^2 + r^2 d \Omega^2,
\end{equation}
where $\bar{M} = M/r_0^L$. 

On the other hand, the Schwarzschild-Tangherlini is
\begin{equation}
   ds^2_{ST} = - \left(1- \frac{\mu}{r^{D-3}}\right) dt^2 +  \left(1- \frac{\mu}{r^{D-3}}\right)^{-1} dr^2 + r^2 d \Omega^2_{D-2},
\end{equation}
where $\mu$ is a parameter related to the black hole mass $M$ by $\mu  = 16\pi M/(D-2)\Omega_{D-2}$, with $\Omega_{D-2} = 2 \pi^{\frac{D-1}{2}}/\Gamma\left(\frac{D-1}{2} \right)$ and $d \Omega^2_{D-2}$ is the metric of the (D-2)-dimensional unit sphere.
\begin{equation}
    d \Omega^2_{D-2}  = d\theta_1^2 + \sin^2 \theta_1 d\theta_2^2+ ... + \prod_{i=1}^{D-3} \sin^2 \theta_i d\theta^2_{D-2}.
\end{equation}


Note that, if we set $D = 4-L$, we obtain
\begin{equation}
    ds^2_{ST} = - \left( 1 - \frac{2M'_L}{r^{1-L}} \right) dt^2 +  \left( 1 - \frac{2M'_L}{r^{1-L}} \right)^{-1}dr^2 + r^2 d\Omega^2_{2-L},
\end{equation}
which is exactly the B-metric, with $M'_L = 4 \pi^{-(1+L)/2}\Gamma\left( \frac{3-L}{2} \right)(2-L)^{-1} M$, except for the last term. Although the parameter $L$ is very small, the two metrics coincide only when $L=0$, but in this case we no longer have Lorentz symmetry breaking (LSB).


We can now verify whether the radial LSB modifies the Schwarzschild black hole thermodynamics. For this purpose, we will use the quantum tunneling formalism.

\section{\textbf{Quantum Tunneling Method}}
\label{tunneling}
Is this section, for illustrative purposes, we review the tunneling approach \cite{Anacleto,Anacleto2} to derive the black hole thermodynamical properties. Besides the Hawking original derivation, many methods were developed in order to derive the black hole thermodynamical properties \cite{Hawking1977,Jacobson,Visser,vagenas2006,Ralf,Parikh,Wilczek}. The tunneling method is a semiclassical approach and its basic idea is to interpret the Hawking radiation as a process of quantum mechanical emission through the black hole horizon. This emission occurs due to the spontaneous creation of particles just inside the black hole horizon. One of the particles tunnels the horizon towards the infinity emerging with positive energy, while the other one with negative energy remains inside the hole and contribute to the mass loss of the black hole. In this sense, it is possible to derive the tunneling probability and associate it with the black hole temperature.

Among the advantages of using the tunneling method, we can cite: it is very useful to incorporate back-reaction effects since it provides a dynamical model of the radiation process; it is related only with the geometry of the spacetime, so it can be applied to a wide variety of spacetimes \cite{Jiang,Angheben,Mann,Wu}; the calculations involved are straightforward. It can be obtained by two ways, namely, the null geodesic method proposed by Parikh and Wilczek \cite{Parikh}, and the Hamilton-Jacobi ansatz originated from the work of Padmanabhan et al. \cite{Pad, Pad2} and used by Angheben et al \cite{Angheben}. However, the two approaches are basically equivalent in many cases. The Hamilton-Jacobi method is more direct and for this reason is the one used in this work \cite{Mann}. In the Hamilton-Jacobi method, we use the so-called WKB approximation in order to obtain the imaginary part of the action. The self-gravitation effects of the particle are not considered.

Near the black hole horizon, we have only the temporal and the radial terms of metric, since the angular part is red-shifted away. The metric becomes 2-dimensional and can be rewritten as

\begin{equation}
ds^{2}=-f(r)dt^{2}+g(r)^{-1}dr^{2}.\label{ds2}
\end{equation}

The Klein-Gordon equation for the field $\phi$ is given by
\begin{equation}
\hslash^{2} g^{\mu\nu}\nabla_{\mu}\nabla_{\nu}\phi-m^{2}\phi=0.
\end{equation}
The last equation with aid of Eq. \eqref{ds2} leads to
\begin{equation}
-\partial_{t}^{2}\phi+\Lambda(r)\partial^{2}_{r}\phi+\frac{1}{2}\partial_{r}\Lambda(r)\partial_{r}\phi-\frac{m^{2}}{\hslash^{2}}f(r)\phi=0,\label{eqphi}
\end{equation} 
where we have defined
\begin{equation}\Lambda(r)\equiv f(r)g(r). 
\end{equation}

Using the so called WKB method \cite{wkb}, we have the following solution for Eq. \eqref{eqphi}
\begin{equation}
\phi(t,r)=\exp\left[-\frac{i}{\hslash}\mathcal{I}(t,r)\right].
\end{equation}

For the lowest order in $\hslash$, we have
\begin{equation} 
(\partial_{t}\mathcal{I})^{2}-\Lambda(r)(\partial_{r}\mathcal{I})^{2}-m^{2}f(r)=0,
\end{equation}
with
\begin{equation}
\mathcal{I}(t,r)=-\omega t+W(r),
\end{equation}
as solution. The explicit form for $W(r)$ is
\begin{equation}
W(r)=\int\frac{dr}{\sqrt{f(r)g(r)}}\sqrt{\omega^{2}-m^{2}f(r)}.\label{eqW1}
\end{equation}

Now we take the approximation of the functions $f(r)$ and $g(r)$ near the event horizon $r_{+}$,
\begin{equation} f(r)=f(r_{+})+f'(r_{+})(r-r_{+})+\cdots, \end{equation}
\begin{equation} g(r)=g(r_{+})+g'(r_{+})(r-r_{+})+\cdots. \end{equation}
The Eq. \eqref{eqW1} then becomes

\begin{equation}
W(r)=\int \frac{dr}{\sqrt{f'(r_{+})g'(r_{+})}}\frac{\sqrt{\omega^{2}-m^{2}f'(r_{+})(r-r_{+})}}{(r-r_{+})},
\end{equation} where the prime denotes derivative with respect to the radial coordinate.

The last integral can be made using the residue theorem, which results in

\begin{equation} W=\frac{2\pi i \omega}{\sqrt{f'(r_{+})g'(r_{+})}}. \end{equation}

The particles tunneling rate is given by $\Gamma\cong\exp\left[-\frac{2}{\hslash}Im\mathcal{I}\right]$, so 
\begin{equation}\label{Gama}
\Gamma
\cong\exp\left[-2Im\mathcal{I}\right]=\exp\left[-\frac{4\pi\omega}{\sqrt{f'(r_{+})g'(r_{+})}}\right].
\end{equation}

Comparing Eq. \eqref{Gama} with the Boltzmann factor, namely $e^{-\omega/T}$, we can obtain the Bekenstein-Hawking temperature, which is given by
\begin{equation} \label{eq:9}
T_{BH}=\frac{\omega}{2Im\mathcal{I}}=\frac{\sqrt{f'(r_{+})g'(r_{+})}}{4\pi}.\end{equation}

From the temperature above, we can obtain the black hole entropy by making use of the
thermodynamical relation $TdS = dM$, which yields to
\begin{equation}
S_{BH} = \int \frac{dM}{T(M)}.
\end{equation}

We can now test whether the tunneling method gives the same results as the Hawking's method. For this purpose, we apply the Hamilton-Jacobi method for a Schwarzschild black hole. In this case, the metric is given by 
\begin{equation} \label{eq:10}
ds^2 = -f(r)dt^{2}+f(r)^{-1}dr^{2},
\end{equation}
where $f(r) = 1 - \frac{2M}{r}$. Using in Eq. \eqref{eq:9} the metric given in Eq. \eqref{eq:10}, we can obtain $T = (8 \pi M)^{-1}$ and $S = 4 \pi M^2$, which are the same results found by Hawking \cite{Hawking}.

Hawking's original derivation gives us
\begin{equation}
T = \frac{\kappa}{2 \pi},
\end{equation}
where $\kappa$ is the surface gravity, given by
\begin{equation}
k^{\mu}\nabla_{\mu} k_{\nu} = \kappa k_{\nu},
\end{equation}
with $k_{\nu}$ being the Killing vector. For a metric given by \eqref{ds2}, we have $k_{\nu} = (1,0,0,0)$ and
\begin{equation}
\kappa = \frac{\sqrt{f'(r_{+})g'(r_{+})}}{2},
\end{equation}
which makes clear that Hawking's method is equivalent to quantum tunneling method.

\section{Thermodynamic Properties for the Bumblebee Model with C-metric} \label{sec:4}
\label{casana}

We first consider the purely radial LSB metric obtained in Ref. \cite{casana}, namely
\begin{equation} \label{eq:20}
ds^{2}=-f(r)dt^{2}+g(r)^{-1}dr^{2}+d\Omega^{2},
\end{equation}
where
\begin{equation} 
f(r)=1-\frac{2M}{r},
\end{equation}

\begin{equation}
g(r)= (1 + \ell)^{-1}\left(1-\frac{2M}{r}\right).
\end{equation}
The surface gravity is given by
\begin{equation}
\kappa_C = \frac{{\kappa_0}}{\sqrt{1+\ell}},
\end{equation}
where $\kappa_0 = 1/4M$ is the Schwarzschild surface gravity, while the temperature is given by
\begin{equation}\label{eq8}
T_{C}= \frac{\sqrt{f'(r_{S})g'(r_{S})}}{4\pi}=\frac{1}{\sqrt{\ell + 1}}\frac{1}{8 \pi  M} = \frac{ T_0}{\sqrt{1 + \ell}},
\end{equation}
where $T_0 = 1/8 \pi M$ is the Hawking temperature for a typical Schwarzschild black hole.

Assuming the approximation $\ell<<1$, the temperature becomes

\begin{equation}
T_{C} \approx \frac{1}{8\pi M}-\frac{\ell}{16\pi M} = \left(1 - \frac{\ell}{2}\right)T_0.
\label{temp1}
\end{equation}
From the right side of Eq. \eqref{temp1}, we can notice that the LSB modified metric given in Eq. \eqref{eq:30} contributes to decrease the temperature of the Schwarzschild black hole.

The entropy is then given by
\begin{equation}
S_{C}=\int\frac{dM}{T(M)}= 4 \pi M^{2} \sqrt{1+ \ell} =\sqrt{1 + \ell}S_0,
\end{equation}
where we used $T$ as given in Eq. \eqref{eq8}, $r_{C} = 2M$ and the Shwarzschild black hole entropy given by $S_0 = A/4 = 4 \pi M^2$, which is provided by the area law. 
This means that, while the black hole temperature obtained by the metric given in Eq. \eqref{eq:30} is smaller than the Schwarzschild one, its entropy is bigger than the usual Scwarzschild entropy. On the other hand, since the black hole surface area is given by $A = 4\pi r_{+}^{2}\sqrt{\ell + 1}$, we can obtain, by using the area law, the same result obtained by the tunneling method. 

At this point, it is worth to investigate the thermal stability of the C-metric, which can be made by analyzing its heat capacity behavior. It is known that Schwarzschild black holes have a negative heat capacity and are always unstable in an infinite bath \cite{hawking2}. This occurs due to the fact that the Schwarzschild temperature decreases as the black hole absorbs mass \cite{wald}. Some Lorentz violating models, however, have nonzero heat capacity which can lead to stable configurations and even to black hole remnants \cite{Ahmed,Ahmed2,Ahmed3,rainbowthermo}. We will consider the heat capacity at constant volume \cite{davies}, namely
\begin{equation}
C_C = T \frac{\partial S}{\partial T} = T \frac{\partial S}{\partial r_S} \left( \frac{\partial T}{\partial r_S} \right)^{-1} = \sqrt{1 + \ell} C_0,
\end{equation}
where $C_0 = - 8 \pi M^2$ is the usual Schwarzschild heat capacity at constant volume. In spite of the term $\sqrt{1 + \ell}$, the heat capacity for the black hole with C-metric has the same features as the Schwarzschild one. For instance, the black hole represented by the C-metric is unstable in spite of the Lorentz symmetry violation \cite{wald}.

Using the same approximation as in Eq. \eqref{temp1}, we can find
\begin{eqnarray}
S_C = \left( 1 + \frac{\ell}{2} \right)S_0,
 \nonumber \\
C_C = \left( 1 + \frac{\ell }{2}\right)C_0,
\end{eqnarray}
which indicates that the dependence of the thermodynamic properties on the LSB parameter is linear.

\section{Thermodynamic Properties for the Bumblebee Model with B-metric}
\label{sec:5}
Now we consider the purely radial LSB metric obtained by Bertolami and P\'aramos \cite{bertolami}, given by
\begin{equation}
ds^2 = - f(r)dt^2 + f(r)^{-1}dr^2 + d\Omega^2,
\end{equation}
where
\begin{equation}
f(r) = 1 - \frac{2M}{r} \frac{r^L}{r_0^L}.
\end{equation}
By performing the same calculation as before, we can determine the temperature and the entropy for this system. As the event horizon radius is now $r_B = (2Mr_0^{-L})^{1/1-L}$, where $r_0$ is a arbitrary distance, we have
\begin{equation} \label{eq:22}
T_{B} = \frac{\sqrt{f'(r_S)g'(r_S)}}{4\pi} = (1 - L)(2Mr_0^{-1})^{-L/(1-L)}T_0.
\end{equation}
This result is the same that Bertolami and P\'aramos obtained by calculating the surface gravity, unless the factor $(1-L)$. Here, we must point out that Bertolami's derivation was not the most accurate since it is used the Schwarzschild surface gravity without LSB. However, a more careful derivation should consider the surface gravity of B-metric, which is given by
\begin{equation}
\kappa_B = (1-L)(2Mr_{0}^{-1})^{-L/(1-L)} \kappa_{0},
\end{equation}
where $\kappa_{0} = 1/4M$ is the Schwarzschild surface gravity. Again, Hawking's method is in agreement with tunneling method.


The next step is to calculate the entropy of this system. We can write the  mass dependence of the temperature as $M^{-1/1-L}$. Therefore,
\begin{equation}
S_{B} = \int \frac{dM}{T(M)} = \frac{2(2M r_0^{-1})^{L/1-L}}{2-L}S_0,
\end{equation}
where $S_0$ is the usual entropy for the Schwarzschild black hole. Using the area law, we find
\begin{equation} \label{eq:23}
S'_B= (2Mr_{0}^{-1})^{\frac{2L}{1-L}} S_{0},
\end{equation}
which differs from the quantum tunneling result. This means that the standard thermodynamics for black holes does not hold for the B-metric. To solve this problem, we need to modify the first law in the following way \cite{zhao, neves}
\begin{equation} \label{eq:24}
T_B dS_B = F(M, r_0, L) dM, 
\end{equation}
where $F$ is a function to be determined, and we have chosen it to be function of $M, r_0$ and $L$ for convenience. In order to determine $F$, we can notice that
\begin{equation}
F(M, r_0, L) = T_B \frac{dS'_B}{dM}.
\end{equation}
We can substitute Eqs. \eqref{eq:22} and \eqref{eq:23} in the last equation and then find
\begin{equation}
F(M, r_0, L) = (2Mr_0^{-1})^{\frac{L}{1-L}}.
\end{equation}
Thus, if we consider the temperature obtained by the tunneling method and the modification in the first law (Eq. \eqref{eq:24}), we can obtain the entropy given by \eqref{eq:23}, which is in agreement with the area law. 

We now calculate the heat capacity for the bumblebee model with B-metric. To do this, we should rewrite the temperature and the entropy in terms of the radius $r_B = (2Mr_0^{-L})^{1/1-L}$ such that
\begin{equation}
T_B = \frac{1-L}{4 \pi} r_B^{-1},
\end{equation}
\begin{equation}
S_B = \pi r_B^2.
\end{equation}
Therefore, the heat capacity for the bumblebee model with B-metric is given by
\begin{equation}
C_B = T \frac{\partial S}{\partial r_B} \left( \frac{\partial T}{\partial r_B} \right)^{-1} = (2Mr_0^{-1})^{\frac{2L}{1-L}} C_0.
\end{equation}
We can see that, as in the case of C-metric, the radial LSB does not modify the main features of the Schwarzschild heat capacity.

Since $L<<1$, the temperature becomes
\begin{eqnarray}
T_B &\approx & \left\lbrace 1 - L[\ln (2M r_0^{-1}) + 1] \right\rbrace T_0,
\end{eqnarray}
while the entropy and the heat capacity become, respectively,
\begin{equation}
S'_B \approx  \left[1 + 2 L \ln (2M r_0^{-1}) \right] S_0,
\end{equation}
and 
\begin{equation}
C_B \approx \left[1 + 2 L \ln (2M r_0^{-1}) \right] C_0.
\end{equation}
As in the C-metric case, we can see that the thermodynamic quantities obtained for the B-metric depend linearly on the LSB parameter, but the dependence on the arbitrary distance $r_0$ is logarithmic. Therefore it is interesting to analyze how these thermodynamic quantities would be influenced by $r_0$ in some specific cases in which the logarithmic dependence is simplified:
\begin{itemize}
\item $r_0 < 2M$: In this case, we have
\begin{equation}
r_{B} = (2 M r_0^{-L})^{1/(1-L)} > 2M > r_0.
\end{equation} 
The arbitrary distance $r_0$ can be interpreted as the distance from the source for which the LSB effects are detected as due to a Yukawa potential \cite{bertolami}. This distance should be higher than the event horizon radius, which does not occur in the present case.

\item $r_0 = 2M$: 
In this case, the arbitrary radius is equal to the event horizon radius $r_{B}= r_0 = 2 M$. Then, we have $\ln (2M r_0^{-1})=0$ and 
\begin{eqnarray}\label{eq:31}
T_B = (1-L)T_0,
\end{eqnarray}
\begin{equation}\label{eq:44}
S'_B \approx S_0,
\end{equation}
\begin{equation}
C_B \approx C_0.
\end{equation}
We can see that the temperature is higher than the usual Schwarzschild temperature. The entropy and the heat capacity, however, are approximately equal to their Schwarzschild correspondents and the horizon radius is equal to the Schwarzschild radius. In this way, the Lorentz symmetry is partially recovered in the limit $r_0 \rightarrow 2M$, but we still have some LSB remnants.

\item $r_0 > 2M$: Now, we have the arbitrary radius $r_0$ is greater than the event horizon radius
\begin{equation}
r_{B} = (2 M r_0^{-L})^{1/(1-L)} < 2M < r_0.
\end{equation}
In this case, $\ln (2M r_0^{-1})<0$, but this condition is not enough for determine whether the temperature and entropy are greater or smaller than the usual ones. However, if we suppose that $2M r_0^{-1}<e^{-1}$, or $r_0 > e(2M)$, where $e = 2,718$ is the Euler number, we have $\ln (2M r_0^{-1})+ 1<0$. Thus, 
\begin{eqnarray}
T_B > T_0.
\end{eqnarray}
In this case, the LSB increases the black hole temperature. When we have $r_0 = e(2M)$ the temperature $T_B$ approaches the Schwarzschild temperature. On the other hand, the entropy and the heat capacity become, as $r_0 \rightarrow e(2M)$,
\begin{equation}
S'_B \approx  (1 - 2 L) S_0,
\end{equation}
and
\begin{equation}
C_B \approx  (1 - 2 L) C_0.
\end{equation}

Under these conditions, we can see that the black hole entropy and heat capacity are decreased by the LSB even though the temperature is approximately equal to the Schwarzschild one.	This case is similar to the last one in the sense that some of the usual results are recovered but we still have LSB remnants.
\end{itemize}

\section{\label{Conclusion}Conclusion}
In this work, we consider black holes thermodynamic properties for black hole scenarios where the Lorentz symmetry is not preserved. We use two different metrics based in the work of Bertolami and Paramos \cite{bertolami} and in the work of Casana et al. \cite{casana}; both metric presented in the cited works are obtained from a radial LSB using the bumblebee formalism. The methodology used was the quantum tunneling and, more specifically, the Hamilton-Jacobi ansatz. We obtained the temperature, entropy and heat capacity for both scenarios.

We showed that the Schwarzschild black hole thermodynamic properties are modified when a Lorentz symmetry violation is taken into account. The temperature change, for example, happens due to the surface gravity modification, which is a consequence of the geometry modification caused by the LSB. The geometry modification is also responsible for changing the black hole surface area and, consequently, the black hole entropy. The temperature and entropy modifications lead to heat capacity modification. However, the black hole instability is not modified by the LSB.

For the C-metric it was verified that the Laws of black hole thermodynamics are not broken and that the LSB could provide corrections to the Schwarzschild thermodynamic quantities. For the B-metric, however, the area law is violated which can be a result of its horizon radius modification. To remedy this issue, we modified the first law of black hole thermodynamics. Subsequently, we discussed situations in which the Schwarzschild thermodynamics could be partially recovered.

\section*{Acknowledgments}

We are grateful to CNPq (Grants No. 305678/2015-9 - RVM, No. 305766/2012-0 - CASA), to CNPq-DAG, and to FUNCAP (PNE-0112-00061.01.00/16 -RVM), for partial financial support. 

\end{document}